\documentclass[prc,twocolumn,superscriptaddress,showpacs,twoside,floatfix]{revtex4}
\usepackage{graphicx}
\usepackage{dcolumn}
\usepackage{bm}

\begin{document}

\title{Proton-proton correlations observed in two-proton decay of  $^{19}$Mg and $^{16}$Ne}

\author{I.~Mukha}
 \affiliation{Universidad de Sevilla, ES-41012 Seville, Spain}
 \affiliation{RRC ``Kurchatov Institute'', RU-123184 Moscow, Russia}
\author{L.~Grigorenko}
 \affiliation{Joint Institute for Nuclear Research, RU-141980 Dubna, Russia}
 \affiliation{Gesellschaft f\"{u}r
Schwerionenforschung,
      D-64291 Darmstadt, Germany}
\author{ K.~S\"{u}mmerer}
 \affiliation{Gesellschaft f\"{u}r
Schwerionenforschung,
      D-64291 Darmstadt, Germany}
\author{L.~Acosta}
 \affiliation{Universidad de Huelva, ES-21071 Huelva, Spain}
\author{M.~A.~G.~Alvarez}
 \affiliation{Universidad de Sevilla, ES-41012 Seville, Spain}
\author{E.~Casarejos}
 \affiliation{Universidade de Santiago de Compostela, ES-15782 Santiago de Compostela, Spain}
\author{A.~Chatillon}
 \affiliation{Gesellschaft f\"{u}r
Schwerionenforschung,
      D-64291 Darmstadt, Germany}
\author{D.~Cortina-Gil}
 \affiliation{Universidade de Santiago de Compostela, ES-15782 Santiago de Compostela, Spain}
\author{J.~Espino}
 \affiliation{Universidad de Sevilla, ES-41012 Seville, Spain}
\author{A.~Fomichev}
 \affiliation{Joint Institute for Nuclear Research, RU-141980 Dubna, Russia}
\author{J.~E.~Garc\'{\i}a-Ramos}
 \affiliation{Universidad de Huelva, ES-21071 Huelva, Spain}
\author{H.~Geissel}
 \affiliation{Gesellschaft f\"{u}r
Schwerionenforschung,
      D-64291 Darmstadt, Germany}
\author{J.~G\'omez-Camacho}
 \affiliation{Universidad de Sevilla, ES-41012 Seville, Spain}
\author{J.~Hofmann}
 \affiliation{Gesellschaft f\"{u}r
Schwerionenforschung,
      D-64291 Darmstadt, Germany}
\author{O.~Kiselev}
  \affiliation{Gesellschaft f\"{u}r
Schwerionenforschung,
      D-64291 Darmstadt, Germany}
\affiliation{Johannes Gutenberg Universit\"at, D-55099 Mainz,
Germany}
\author{A.~Korsheninnikov}
\affiliation{RRC ``Kurchatov Institute'', RU-123184 Moscow, Russia}
\author{N.~Kurz}
  \affiliation{Gesellschaft f\"{u}r
Schwerionenforschung,
      D-64291 Darmstadt, Germany}
\author{Yu.~Litvinov}
  \affiliation{Gesellschaft f\"{u}r
Schwerionenforschung,
      D-64291 Darmstadt, Germany}
\author{I.~Martel}
 \affiliation{Universidad de Huelva, ES-21071 Huelva, Spain}
\author{C.~Nociforo}
  \affiliation{Gesellschaft f\"{u}r
Schwerionenforschung,
      D-64291 Darmstadt, Germany}
\author{W.~Ott}
 \affiliation{Gesellschaft f\"{u}r
Schwerionenforschung,
      D-64291 Darmstadt, Germany}
\author{M.~Pf\"utzner}
 \affiliation{IEP, Warsaw University, PL-00681 Warszawa, Poland}
\author{C.~Rodr\'{\i}guez-Tajes}
 \affiliation{Universidade de Santiago de Compostela, ES-15782 Santiago de Compostela, Spain}
\author{E.~Roeckl}
  \affiliation{Gesellschaft f\"{u}r
Schwerionenforschung,
      D-64291 Darmstadt, Germany}
\author{M.~Stanoiu}
  \affiliation{Gesellschaft f\"{u}r
Schwerionenforschung,
      D-64291 Darmstadt, Germany}
\affiliation{IFIN-HH, P.~O.~BOX MG-6, Bucharest, Romania}
\author{H.~Weick}
 \affiliation{Gesellschaft f\"{u}r
Schwerionenforschung,
      D-64291 Darmstadt, Germany}
\author{P.~J.~Woods}
 \affiliation{University of Edinburgh, EH1 1HT Edinburgh, UK}

\begin{abstract}

 Proton-proton correlations were observed
for the two-proton decays of the  ground states of $^{19}$Mg and $^{16}$Ne.
The trajectories of the respective decay products, $^{17}$Ne+p+p
and $^{14}$O+p+p, were measured by using a tracking technique with
microstrip detectors. These data were used to reconstruct the
angular correlations of fragments projected on { planes
transverse} to the precursor momenta. The measured three-particle
correlations reflect a genuine three-body decay mechanism and
allowed us to obtain spectroscopic information on the precursors
with valence protons in the $sd$ shell.

\end{abstract}


 \maketitle


The recently discovered  two-proton (2p) radioactivity is  a
specific type of genuine three-particle nuclear decays. It occurs
when a resonance  in any pair of fragments is located at higher
energies than in the initial three-body (p+p+``core'') nucleus, and
thus  simultaneous emission of { two} protons is the only decay
channel. Three-body systems have more degrees of freedom in
comparison with two-body systems, hence additional observables
appear.
 In the case of 2p emission, the energy spectra of single
protons become continuous, and proton-proton (p--p) correlations
are available, which makes them a prospective probe for nuclear
structure or/and { the} decay mechanism. For example, the first
p--p correlations observed in the 2p radioactivity of $^{94m}$Ag
have revealed strong proton yields either in the same or opposite
directions which called for a theory of 2p emission from deformed
nuclei \cite{mukh06}.
Two-proton emission can also occur from short-lived nuclear
resonances or excited states (see, { e.g., }
\cite{boch89,o12,o14}). Though { in this case} the mechanism of
2p emission may depend on the reaction populating the parent
state, such nuclei { can be easily studied in-flight}.
{ E.g., the cases of $^{6}$Be \cite{boch89,dan87} and $^{16}$Ne
\cite{korsh_ne16}were studied by analyzing their p--p correlations
in the framework of a three-body partial-wave analysis developed
for three-particle decays of light nuclei}.
In particular, the study of $^6$Be revealed the existence of
three-particle p+p+$\alpha$ correlations \cite{boch89} which
matched the three-body components found { theoretically} in the
$p$-shell structure of  $^6$Be  \cite{thomp00}. Very recently,
p--p correlations were also observed in 2p radioactivity of
$^{45}$Fe \cite{giov07,mier07} where both the { lifetime and
p--p correlations were found to reflect the} structure of
$pf$-shell 2p precursors \cite{mier07}. Such a way { of obtaining
spectroscopic information is a} novel feature compared to studies
of two-particle decays.

{ In the present paper, we study for the first time} the p--p
correlations in $sd$ shell nuclei { via} examples of the 2p
decays of $^{19}$Mg and $^{16}$Ne. These nuclei with very
different half { lives}  ($T_{1/2} \!
 \approx  \! 4 \! \cdot \! 10^{-9}$ s \cite{mukh_mg19}
 and $T_{1/2} \!  \approx  \! 4 \! \cdot \! 10^{-19}$ s \cite{kek_ne16},
 respectively) and presumably different
spectroscopic properties may serve as { reference cases
illuminating the nuclear structure of other possible 2p emitters
with $sd$-wave} configuration.

\begin{figure}[t!]

\includegraphics[width=0.48\textwidth,angle=0.]{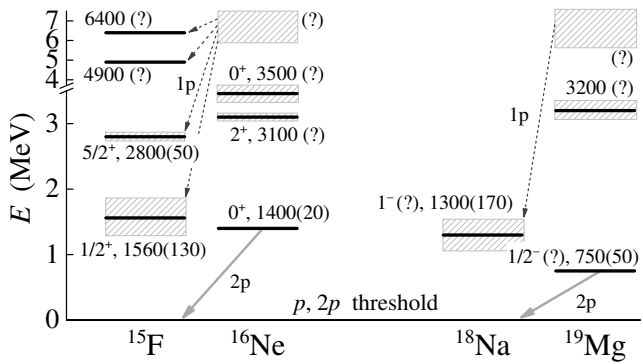}

      \caption{\label{fig0}
States observed in $^{16}$Ne, $^{19}$Mg  and the corresponding
intermediate systems $^{15}$F, $^{18}$Na.  Decay energies (in
keV) are given relative to the respective p and 2p thresholds.
Most values  have been taken from the literature
(Refs.~\cite{mukh_mg19,kek_ne16,ne16,ne16_sc,f15_pet,f15_gol,18Na}),
those in bold print are from the present work.}

      \end{figure}

The decay properties of the $^{16}$Ne and $^{19}$Mg ground states
and the related resonances in $^{15}$F and $^{18}$Na are shown in
Fig.~\ref{fig0} which compiles the data from
Refs.~\cite{kek_ne16,ne16,ne16_sc,f15_pet,f15_gol,18Na,mukh_mg19} and this work.
The ground states of both isotopes decay only { by
simultaneous} 2p emission while their excited states are open for
sequential 1p decays { via intermediate} unbound states in
$^{15}$F and $^{18}$Na.

The  quantum-mechanical theory of 2p radioactivity which uses
a three-body  model \cite{grig00,grig01,grig03},  predicts the
p--p correlations { to be} strongly influenced by nuclear
structure together with
 Coulomb and three-body centrifugal barriers.
In particular,  the newly discovered 2p-radioactivity of $^{19}$Mg
\cite{mukh_mg19} was predicted to be characterized by p--p
correlations which reflect  {the $sd$} configurations of the valence
protons
\cite{grig03a}. A similar effect is found in $^{16}$Ne, where {
the} $s$-wave configuration was predicted to dominate, contrary to
its mirror $^{16}$C, thus breaking isospin symmetry
\cite{grig_ne16}. { A} complementary approach in describing 2p
decays is { the} mechanism of sequential emission of protons
via an intermediate state (see e.g., \cite{lane}). It includes
also the traditional quasi-classical  di-proton model with
emission of a $^2$He cluster, { assuming} extremely strong p--p
correlations \cite{gold60,baz72}. The predictions of these models
differ dramatically { with respect to} the p--p correlations,
suggesting them as { a} sensitive probe of the 2p-decay
mechanism
 (see the detailed predictions below).

{ Our experiment to investigate 2p-emission from $^{19}$Mg and
$^{16}$Ne} was performed by using a 591{\it A} MeV beam of
$^{24}$Mg accelerated by the SIS facility at GSI, Darmstadt. The
radioactive beams of $^{20}$Mg and $^{17}$Ne were produced at the
Projectile-Fragment Separator (FRS) \cite{frs} with average
intensities of 400 and 800 ions s$^{-1}$ and energies of ~450{\it
A} and ~410{\it A} MeV, respectively. The secondary {
1-n-removal} reactions ($^{20}$Mg, $^{19}$Mg) and ($^{17}$Ne,
$^{16}$Ne) occurred at the mid-plane of FRS in a secondary 2
g/cm$^{2}$ $^{9}$Be target. Special magnetic optics settings were
applied, the first FRS half being tuned in an achromatic mode
using a wedge-shaped degrader, while its second half was set for
identification of the heavy ions (HI)
with high acceptance in angle and momentum.

{ A sketch of the experimental set-up at the FRS
midplane has been shown in Fig.1 of Ref.~\cite{mukh_mg19} and
explained in detail there. A microstrip detector array
\cite{si_proposal} consisting of 4 large-area (7x4 cm$^{2}$)
double-sided silicon detectors (with a pitch of 0.1 mm)was
positioned downstream of the secondary target.}
{ This array was} used to measure energy loss and position of
coincident hits of two protons and a heavy fragment, thus {
allowing us to} reconstruct all decay-product trajectories and derive
the coordinates of the reaction vertex and the angular p--p and
proton-HI correlations.
 The conditions { to select} the true HI+p+p events were: (i) { a minimal distance
 between the proton and heavy ion trajectories of less than 150
 $\mu$m},
 and (ii) { a difference between the two longitudinal coordinates of the vertices defined by
 two p--HI pairs (taken from the same HI+p+p event)
 within the range} defined by the experimental uncertainty of 0.3--1 mm
 depending on detection angle.
 The achieved angular resolution in tracking { the} fragments was  $\sim$1 mrad.
More details concerning the detector performance and tracking procedure are given in
\cite{mukh_mg19,mihai,mukh_ssd,readout}.
{ Another position-sensitive silicon detector and a multi-wire}
chamber were used upstream of the target for tracking the
$^{20}$Mg($^{17}$Ne) projectiles. The heavy  2p-out residuals ($^{17}$Ne and
$^{14}$O) were unambiguously identified by their time of flight,
magnetic rigidity, and  energy loss measured with the
position-sensitive scintillator detectors at the second half of
FRS.

      \begin{figure}[t!]

\hspace{-0.3 cm}
\includegraphics[width=0.49\textwidth,angle=0.]{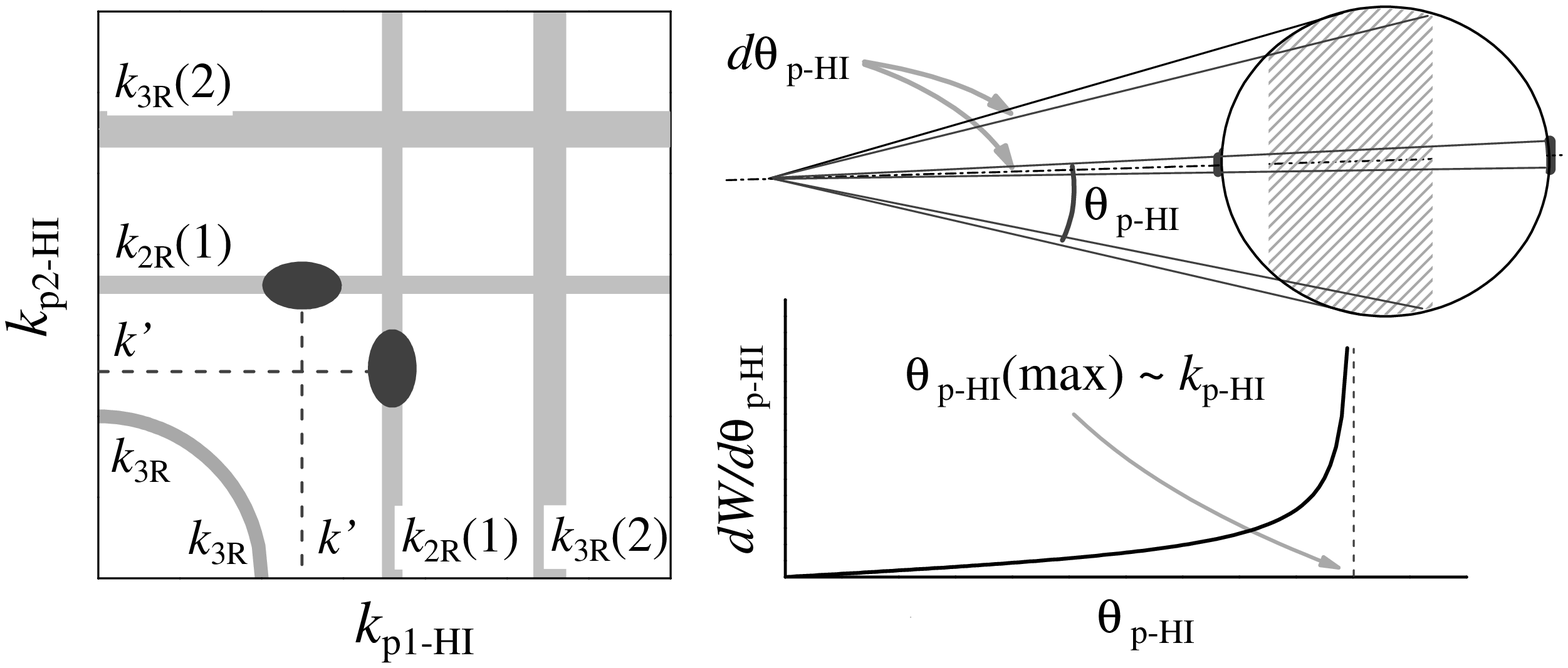}

      \caption{\label{fig2a}
Left panel: { Cartoon} of  momentum correlations
$k_{p1-\!HI}-k_{p2-\!HI}$  expected for a direct { three-body
decay} (the grey area { labeled} $k_{3R}$)  { and
sequential 2p decay (grey boxes with black peaks labeled
$k_{2R}$).} Upper-right panel: A sketch of the kinematical
enhancement of an angular p--HI correlation at the maximum
possible angle for { a} given momentum between the decay
products. Lower-right panel: the corresponding angular p--HI
distribution.}

      \end{figure}
%

%
%
     \begin{figure*}[tbh]
\mbox{\hspace{-0.2cm}
\includegraphics[width=0.29\textwidth,angle=0.]{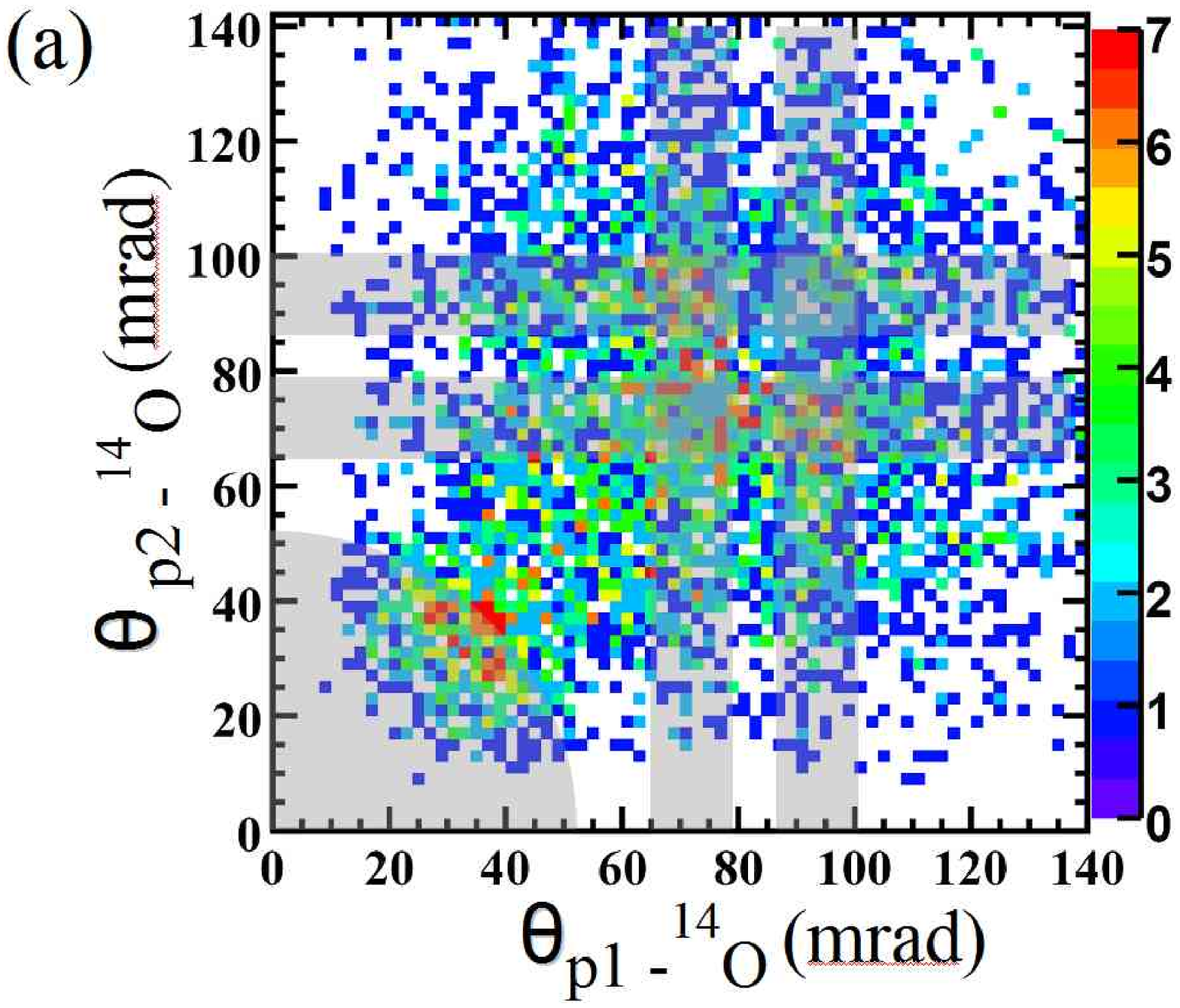}\hspace{0.2cm}
\includegraphics[width=0.34\textwidth,angle=0.]{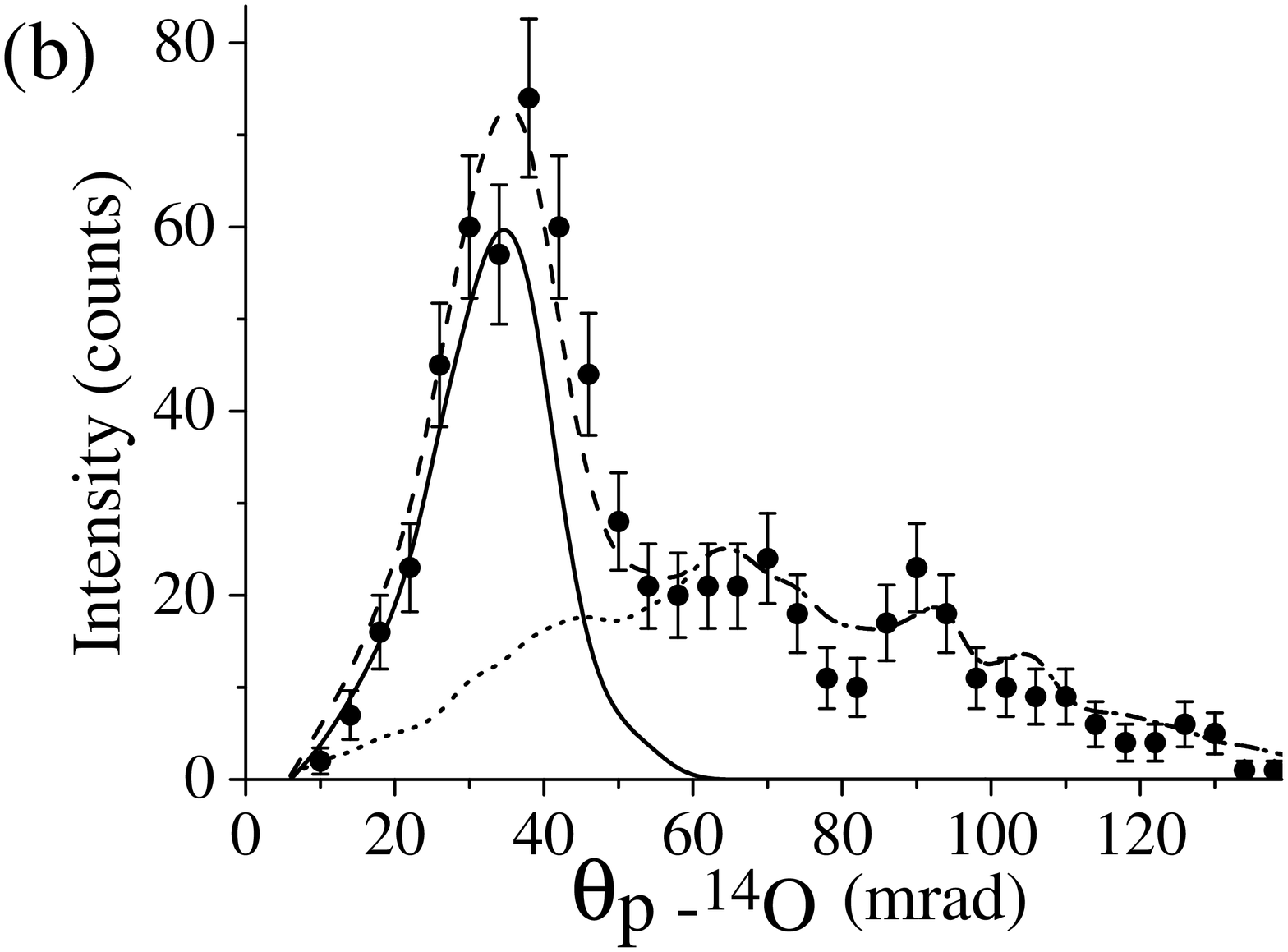}\hspace{-0.70cm}
\includegraphics[width=0.38\textwidth,angle=0.]{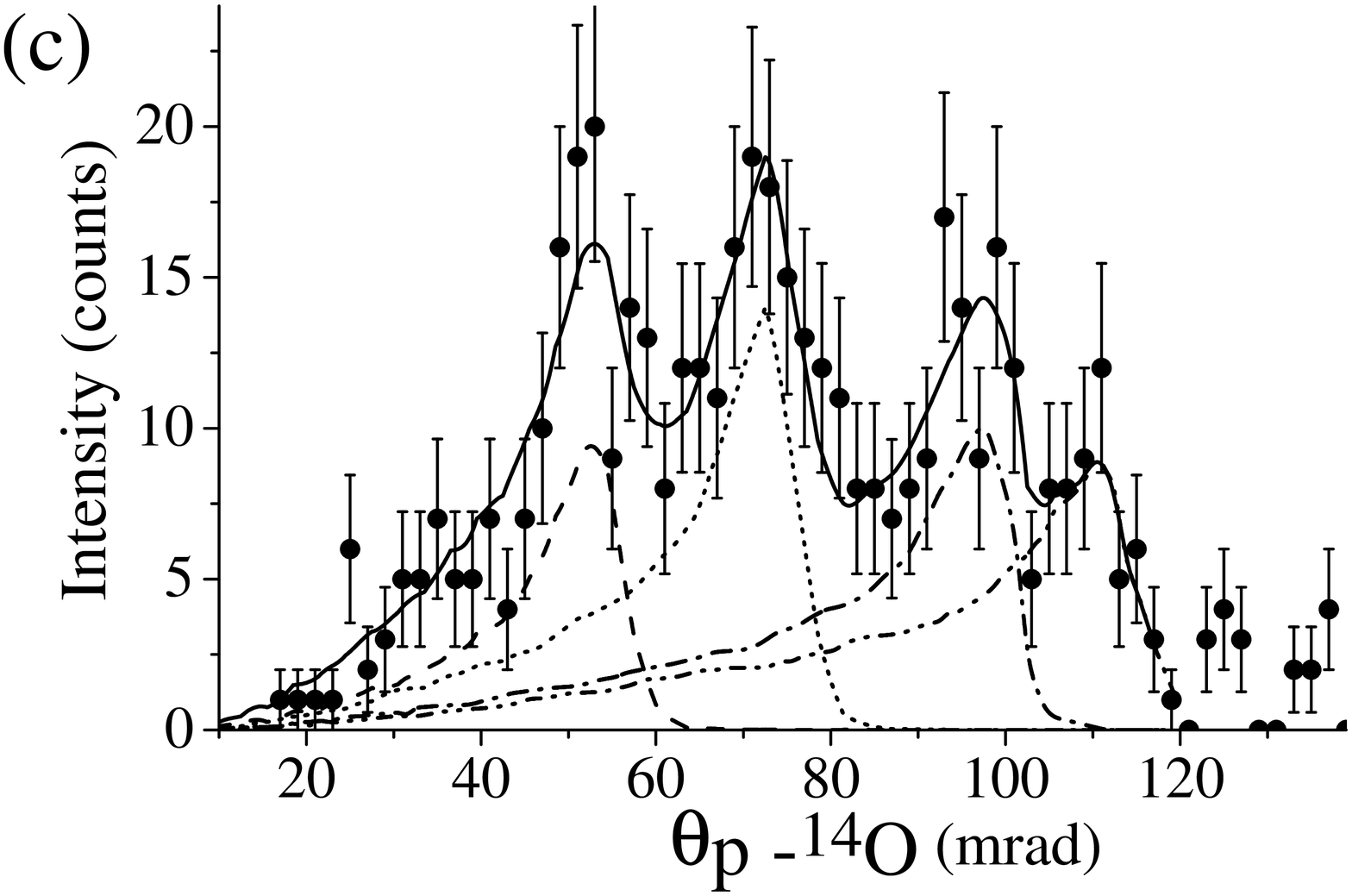}
}

\caption{
      \label{fig2}
(a)  Angular (p$_1$--$^{14}$O)--(p$_2$--$^{14}$O) correlations
  obtained from the  measured $^{14}$O+p+p
  events  ({ colored} boxes with  scale shown on right).
  The grey areas indicate simultaneous and sequential 2p emission in analogy to those shown in Fig.~{\protect\ref{fig2a}}.
(b) Angular p--$^{14}$O distribution
  (full circles with statistical uncertainties) obtained from the data shown { in panel} (a)
  by selecting the other proton angle $\theta_{p2-\!\rm{O}}$ { within the range} from 0 to 45
  mrad, which corresponds to  the ground state of $^{16}$Ne. The solid curve represents the  Monte Carlo
    simulation of the detector response for
    $^{16}$Ne$_{g.s.}\!\!\!\rightarrow^{14}$O+p+p with
    { a} 2p-decay energy of 1.4(1) MeV {\protect \cite{ne16}}.
    The dashed  line is a sum fit to the data. The dotted curve
    indicates the background as explained in the text.
(c) Angular p--$^{14}$O distribution obtained from the data shown
{ in} (a) by selecting the other proton angle from 120 to 150
mrad, which corresponds to p--$^{14}$O final-state interactions
due to the ground and excited states of $^{15}$F. The dashed and
dotted curves are the Monte Carlo simulations of the known 1p
decays of the ground and first excited states of $^{15}$F with the
1p-decay energies of 1.56(13) and 2.85(4) MeV, respectively
{\protect\cite{ne16_sc}}. The { dash-dotted and
dash-dot-dotted} curves show the result of fits to the p--$^{14}$O
correlation, { indicating unknown} excited states in $^{15}$F
with 1p-decay { energies} of 4.9(2) and 6.4(2) MeV,
respectively. The solid line is the sum fit. }
      \end{figure*}
%


The 2p decays of the ground states of $^{19}$Mg and $^{16}$Ne were
identified by using angular correlations between the single
protons and { their} respective cores $^{17}$Ne { and}
$^{14}$O which allowed measuring the 2p-decay energies.
This is analogous to identifying a reaction  { via the scatter
plot illustrated} in Fig.~\ref{fig2a} (left panel).
In general, 2p-decay  may proceed via either sequential
or direct decay mechanisms. The first case can be described as two
consecutive 1p decays, with the p--HI spectra reflecting the
respective p--HI resonances \cite{lane}. { Their
(p$_1$--HI)--(p$_2$--HI) scatter plot should display the sequential and direct decay events
in the respective kinematical areas marked in Fig.~\ref{fig2a} by
the relative momenta $k_{2R}$}. In the second case, the
simultaneously emitted protons are likely to share the 2p-decay energy evenly, with both
p--HI spectra being identical and peaked at $E/2$ \cite{baz72,grig00}.
{ In this case, the area marked $k_{3R}$ in Fig.~\ref{fig2a}
should be populated, along the arc area with the root-mean-squared
proton momentum being constant}. Sequential proton emission from a
single 2p-parent state via narrow p--HI resonances should yield
double peaks while 2p de-excitation of continuum parent states
with p--HI final state interactions should reveal ``slices'' as
shown in Fig.~\ref{fig2a}.

Similar structures can be found in the angular
$\theta_{p1-\!HI}-\theta_{p2-\!HI}$ correlations due to the
following reason. Because of a strong { kinematic} focusing at
intermediate energies, the 1p decay  leads to a characteristic
angular p--HI correlation when the proton ejected isotropically in the 1p-precursor frame
is emitted within a
narrow cone around the HI with the maximum intensity around the
largest possible angle (Fig.~\ref{fig2a}, right panel). The  p--HI
angles reflect the transverse proton momentum relative to the HI
one, and { are therefore} correlated with the precursor's decay
energy. Thus sequential 2p decays from excited states in parent
nuclei { should be mostly located in  peaks with tails along the respective slices in} the angular
$\theta_{p1-HI}-\theta_{p2-HI}$ correlations, in analogy to those
sketched in Fig.~\ref{fig2a} (left panel). In direct 2p decays,
the single-proton energy spectrum always { exhibits} a
relatively narrow peak centered close to half of the 2p-decay
energy; such energy distribution is a stable feature of this decay mechanism \cite{gold60,grig03a}.
Correspondingly, a bump should appear in the angular correlations
in the same way as it should appear along the arc marked
by $k_{3R}$ in the scatter plot in Fig.~\ref{fig2a}. { This
correspondence between angular and momentum correlations has been
used to derive the 2p-decay energy of $^{19}$Mg} (see Fig.~2,4  in
\cite{mukh_mg19} and  the respective discussions).

{ For the 2p-decay of $^{16}$Ne, the} angular correlations of
each coincident proton with respect to  the $^{14}$O momentum,
$\theta_{p1-\!\rm{O}}-\theta_{p2-\!\rm{O}}$, derived from the
measured $^{14}$O+p+p coincidence events, are shown in
Fig.~\ref{fig2}(a). The events with the { smallest} angles fall
into a distinct cluster around $\theta_{p-\!\rm{O}}$=35 mrad while
most of the other events are { located in the slices centered}
around 70 and 95 mrad. These two groups can be attributed to the direct
2p decay from the $^{16}$Ne ground state and to the sequential
emission of protons from excited states in $^{16}$Ne via the
$^{15}$F ground-state, respectively.  We shall refer to these
events as the ``ground state'' and ``excited state'' events,
respectively. The latter group includes also events resulting from
the fragmentation reaction $^{17}$Ne$\rightarrow^{14}$O+p+p+n.

To disentangle the ``ground state'' from the ``excited state''
events, we made a slice projection from the measured
(p$_1$--$^{14}$O)--(p$_2$--$^{14}$O) correlations in
Fig.~\ref{fig2}(a) by selecting the angle of one of the protons
within the range 0--45 mrad, where the 2p decay of the $^{16}$Ne
ground state is expected to show up. Figure \ref{fig2}(b) displays
the angular correlations  $\theta_{p1-\!\rm{O}}$ corresponding to
the ``ground state'' gate in { the other} pair
$\theta_{p2-\!\rm{O}}$. The peak around { 35} mrad (the
suggested ``ground state'') dominates the spectrum, { whereas
few} correlations { can be} seen between a proton from the
``ground state'' and another proton at larger angles. This means
that the two protons from the ``ground-state'' are correlated,  i.e.\ this peak can  be explained
by an emission of protons from the ground state in $^{16}$Ne.

{ For a more quantitative analysis,} the data are compared to a
Monte Carlo simulation of the response of our setup to the direct
2p-decay $^{16}$Ne$\rightarrow^{14}$O+p+p  with the known 2p-decay
energy of  1.4(1) MeV \cite{ne16} by using the GEANT program \cite{GEANT}.
The simulations took into account the above-mentioned experimental
accuracies in tracking the fragments and in { reconstructing
the vertices}, trajectory angles etc. The normalized simulation
reproduces the data in the low-angle peak { very well}.
The contribution from the ``tail'' of the higher states to the ground-state peak
amounts to about $20 \%$. The shape of this distribution is assumed to have the
same shape as the
$\theta_{p1-\!\rm{O}}$ distribution selected within the
$\theta_{p2-\!\rm{O}}$ range just outside the ground-state
region, from 48 to 160 mrad (the dotted curve).

Figure \ref{fig2}(c) displays an example of an angular p--$^{14}$O
distribution for ``excited states'' obtained from
Fig.~\ref{fig2}(a)  by selecting the angular range of the other
proton from 120 to 150 mrad, which corresponds to p--$^{14}$O
final state interactions due to the ground and excited states of
$^{15}$F. The Monte Carlo simulation of known one-proton decays
$^{15}$F$\!\rightarrow^{14}$O+p of the ground and first excited
states { in} $^{15}$F with the 1p-decay energies of 1.56(13) and
2.85(4) MeV \cite{ne16_sc}, respectively, { reproduces well}
the two lowest-angle peaks. The two higher-lying peaks indicate 1p
decays { of unknown} excited states in $^{15}$F { with
derived} 1p-decay { energies} of  4.9(2) and 6.4(2) MeV. The
excited states in $^{19}$Mg, $^{16}$Ne, $^{18}$Na and $^{15}$F
will be addressed elsewhere.

%
      \begin{figure}[t!]

\centerline{\includegraphics[width=0.48\textwidth]{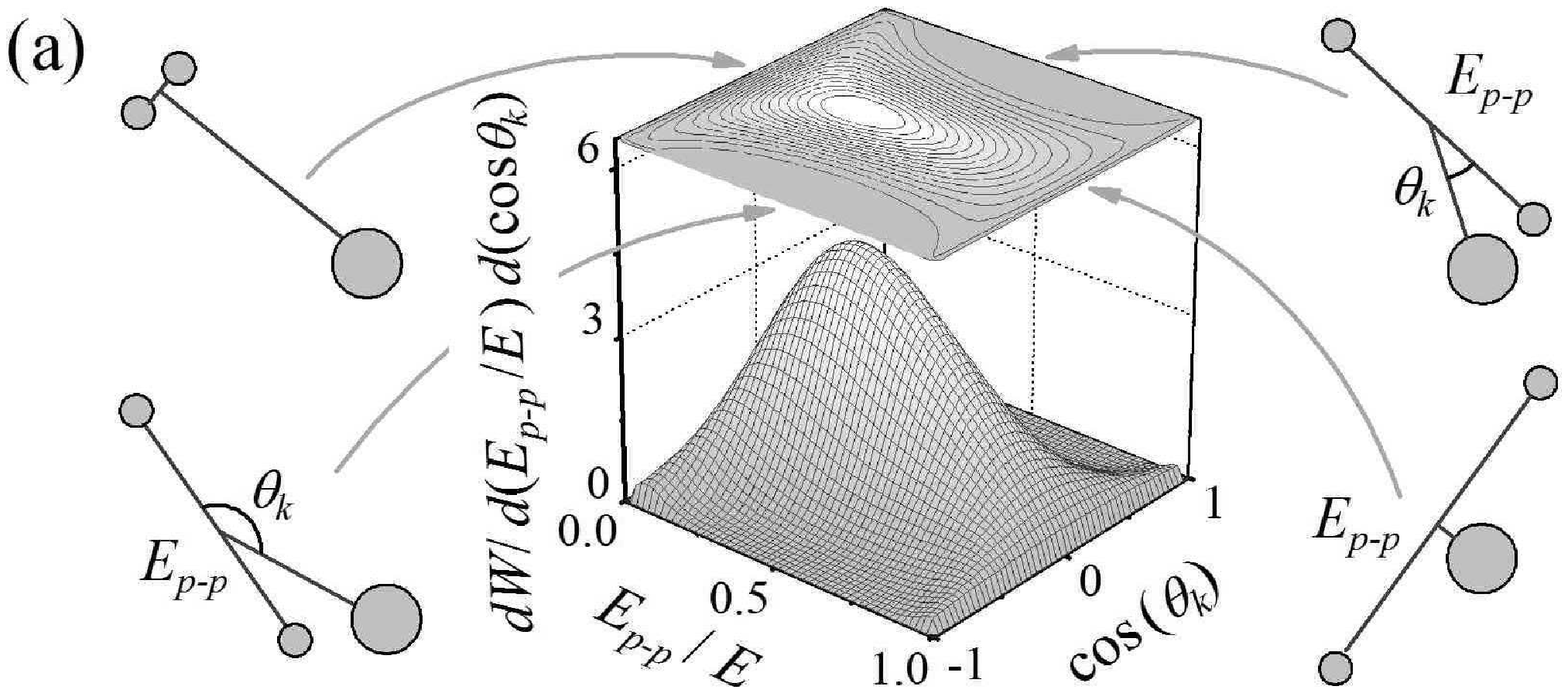}}
\includegraphics[width=0.258\textwidth]{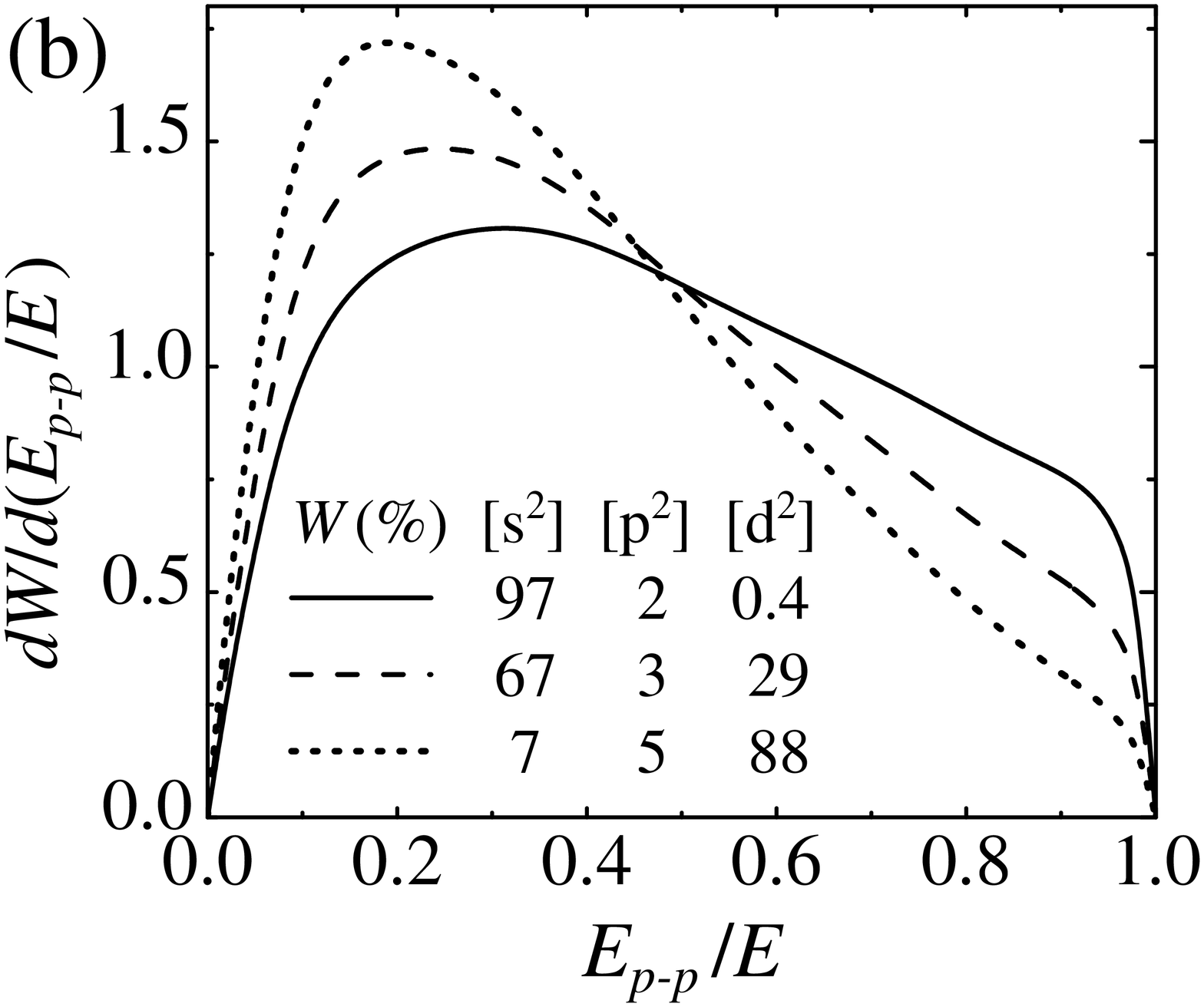}
\includegraphics[width=0.218\textwidth]{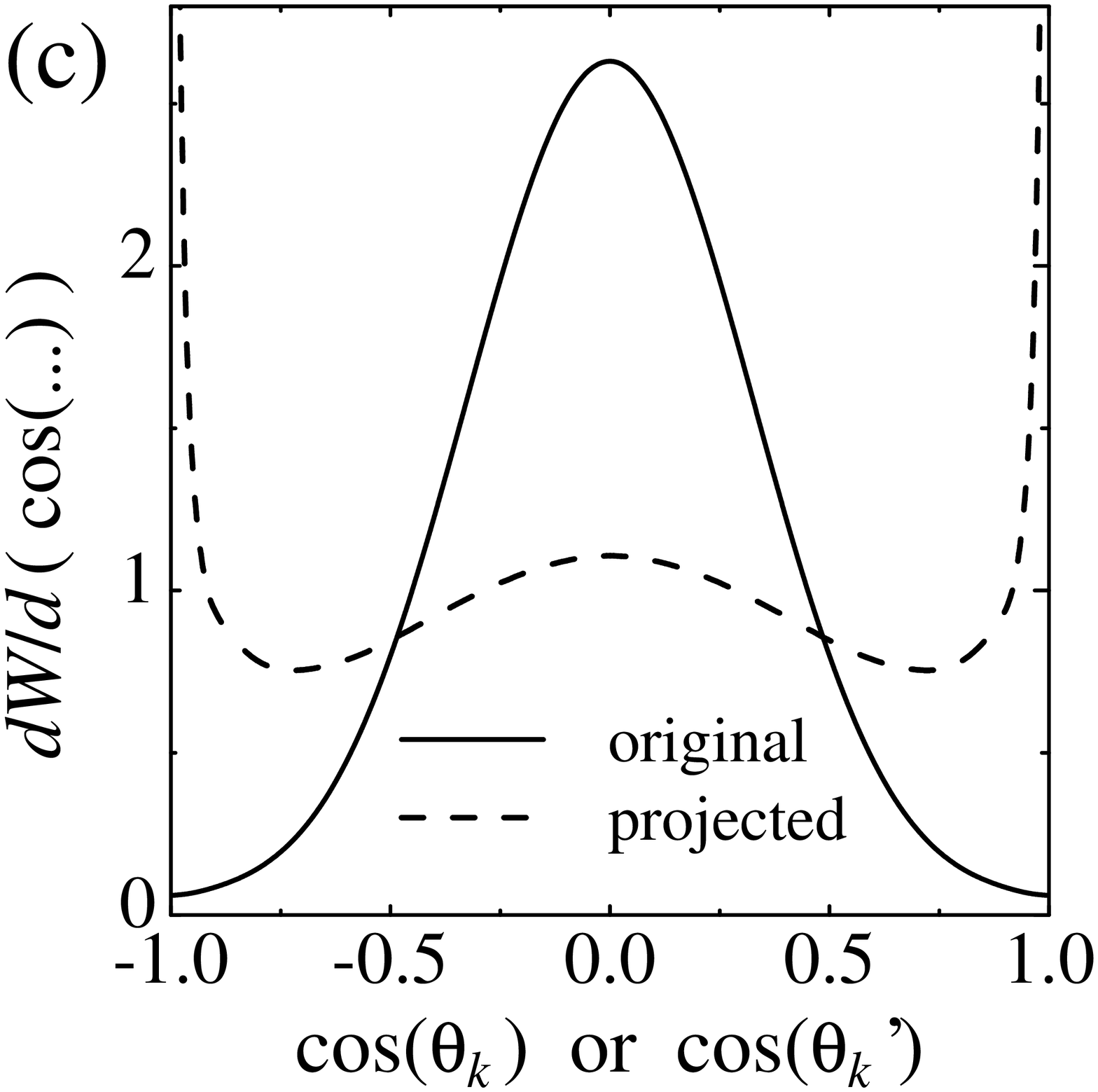}

\caption{
      \label{fig1a}
(a) Three-body correlations predicted by the three-body model
\protect\cite{grig03a} { for the 2p-decay of $^{19}$Mg, plotted
as a function of the relative energy between two protons,
$E_{p-p}$/$E$, and the angle $\theta_k$ between the relative
momenta of the fragments as illustrated in the figure. Extreme
cases of p--p correlations are sketched and related to the
corresponding areas in the correlation plot}. (b) The p--p energy
spectra from { the} 2p-decay of $^{19}$Mg calculated for
different weights $W$ of $s$-$p$-$d$ shell configurations in
$^{19}$Mg. (c) Typical intensity distributions plotted as a
function of $\theta_k$ (see Fig.~{\protect\ref{fig1a}}(a)) in the
rest frame of the { 2p-precursor} (solid curve) and  its analog
in the lab system $\theta'_k$ projected on the { transverse
detector} plane (dashed curve).  }
      \end{figure}
%

%
We turn now to the discussion of angular  p--p correlations
following 2p decays. When the spin degrees of freedom are
neglected and the total decay energy $E$ can be considered as
fixed, the three-body correlations are completely described by
two variables. A convenient choice is an { energy-distribution}
parameter $E_{p-p}/E$ ($E_{p-p}$ is relative energy between two
protons) and an angle $\theta_k$ between the relative momenta of
the fragments. { Fig.\ref{fig1a}} shows such distributions
predicted by the three-body model for { the 2p-decay} of
$^{19}$Mg \cite{grig03,grig03a}. The three-body model {
predicts a} distinctive correlation pattern which features {
an} enhancement at small $E_{p-p}$ due { to final-state
interaction} and { a} suppression in the regions of strong
Coulomb repulsion ($E_{p-p}/E \sim 0.5$, $\cos(\theta_k) \sim
\pm 1$). The predicted energy distributions { are sensitive} to
the structure of the precursor [Fig.~\ref{fig1a}(b)]. Similar
predictions are available for $^{16}$Ne \cite{grig_ne16}.
%

%
      \begin{figure}[ht]

\mbox{\hspace{-0.1 cm}
\includegraphics[width=0.24\textwidth,angle=0.]{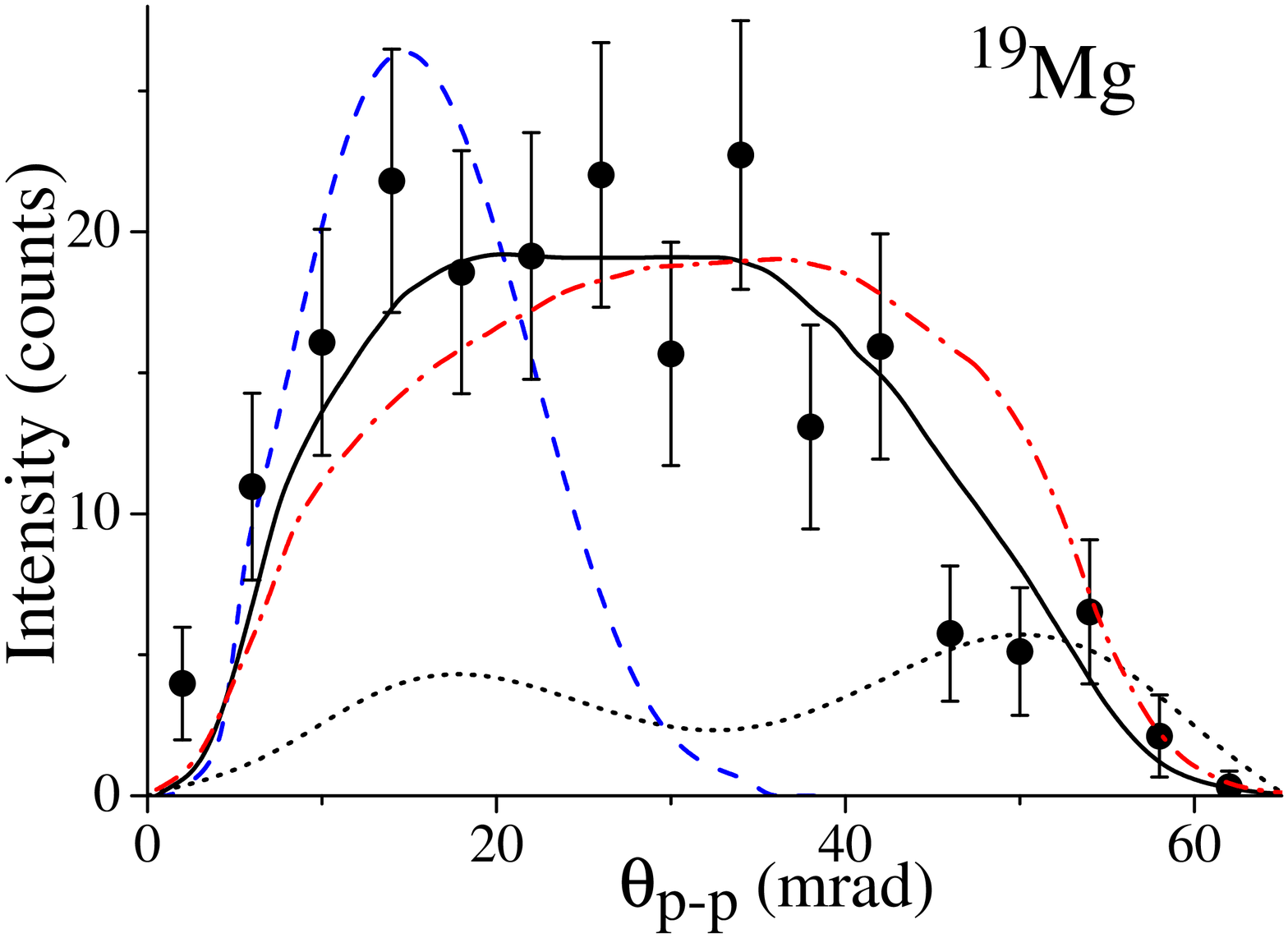}\hspace{-0.2 cm} 
\includegraphics[width=0.24\textwidth,angle=0.]{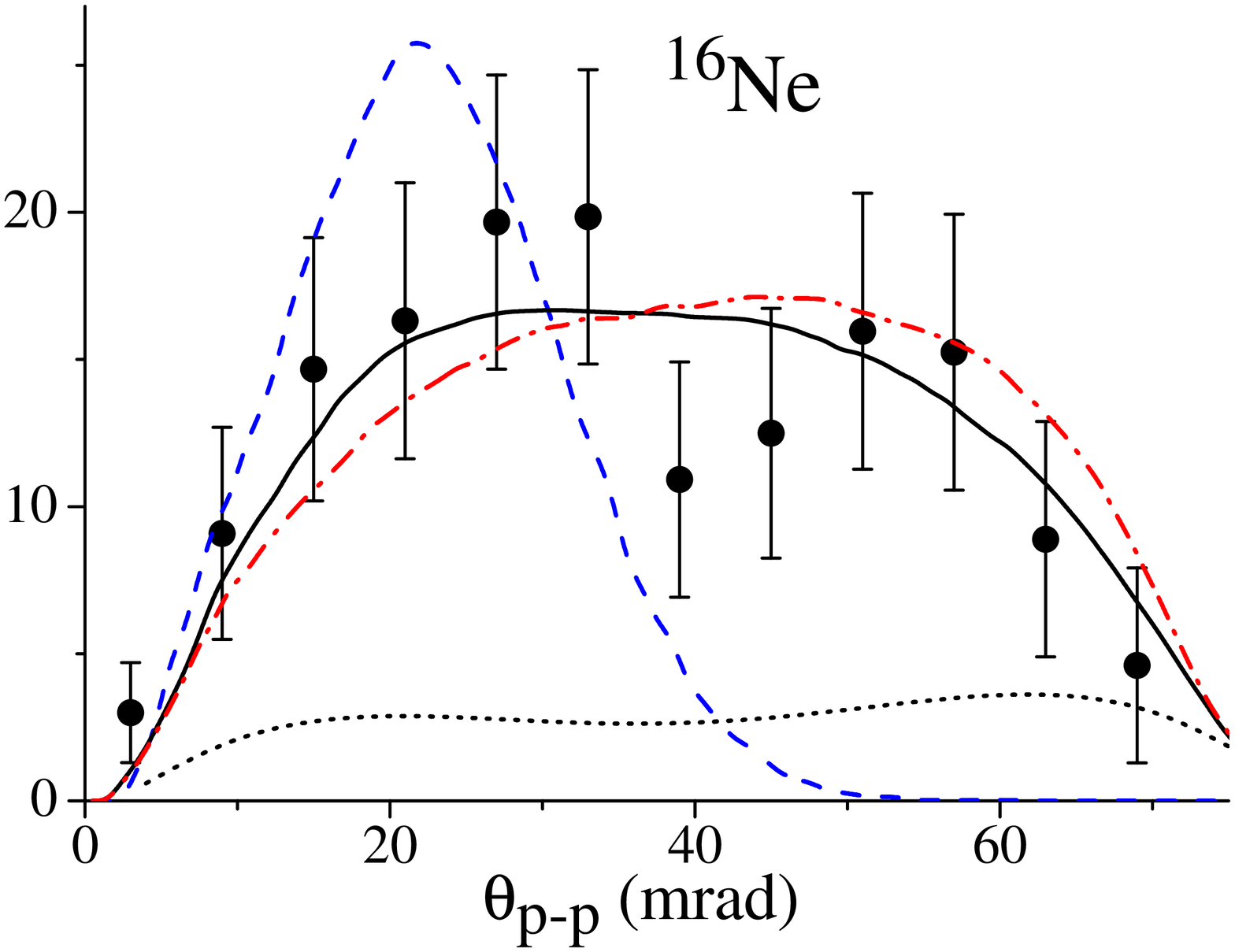} 
}
\vspace{-0.2 cm} \caption{
      \label{fig4}
 Angular p--p correlations from the 2p-decays of the ground states of
 $^{19}$Mg { (left)}
 and $^{16}$Ne { (right)})
 obtained from the  measured $^{14}$O+p+p and $^{17}$Ne+p+p
 { events, respectively,} by selecting both protons from the respective p--$^{14}$O($^{17}$Ne) angular ranges.
 The solid curves show the fits to the data, obtained by using the three-body model calculations (54\% and 88\% of  $d$-wave
 { configuration} in $^{16}$Ne and $^{19}$Mg, respectively.)
 The dotted curves show the background
 contributions estimated as described in the text. The dashed curves are the diproton model predictions, and the dash-dotted curves are the phase-space
 simulations of the 2p-decays { (isotropic proton emission in the 2p-precursor's rest frame)}.
}
      \end{figure}
%

In our experiment, we were able to measure the opening angle
$\theta_{p-p}$ between protons whose distribution reflects the
$E_{p-p}$ correlations. Fig.\ref{fig4} { shows} the
experimental angular p--p distributions obtained from triple
$^{14}$O+p+p and $^{17}$Ne+p+p events gated by the conditions that
both protons originate from the { ``ground states'' of
$^{16}$Ne and $^{19}$Mg}. These gates { were} inferred from the
respective angular p--HI correlations as discussed above.
The events { representing} the ``ground state'' 2p decay
actually contain { contributions from the ground state and
background contributions from both, excited states of the parent
nucleus and fragmentation reactions}. Therefore, using the
angular-correlation data discussed above, we empirically evaluated
the shapes of the background components by projecting triple
events with the p--HI gates shifted { away from the ``ground
state'' region towards larger angles}. The resulting p--p
background contributions shown by the dotted curves in
Fig.~\ref{fig4} { constitute} about 20 \% of all p--p
correlation data for $^{16}$Ne and 25 \% for $^{19}$Mg (see
Fig.~\ref{fig2}(b) and Fig.\ 4 (c) in \cite{mukh_mg19}); {
they} were subtracted from the original p--p { correlations}.
As one can see { in Fig.~\ref{fig4}}, the predictions {
following from} the assumption of a diproton { emission fail to
describe} both the $^{16}$Ne and $^{19}$Mg data while the
three-body model reproduces { the} shapes of both
distributions. In the $^{19}$Mg case, the best description is
obtained with the $d$-wave configuration dominating. The $^{16}$Ne
data give evidence for { nearly equal $s$- and $d$-wave
components}.

%
%
      \begin{figure}[thb]

\mbox{\hspace{-0.1 cm}
\includegraphics[width=0.25\textwidth,angle=0.]{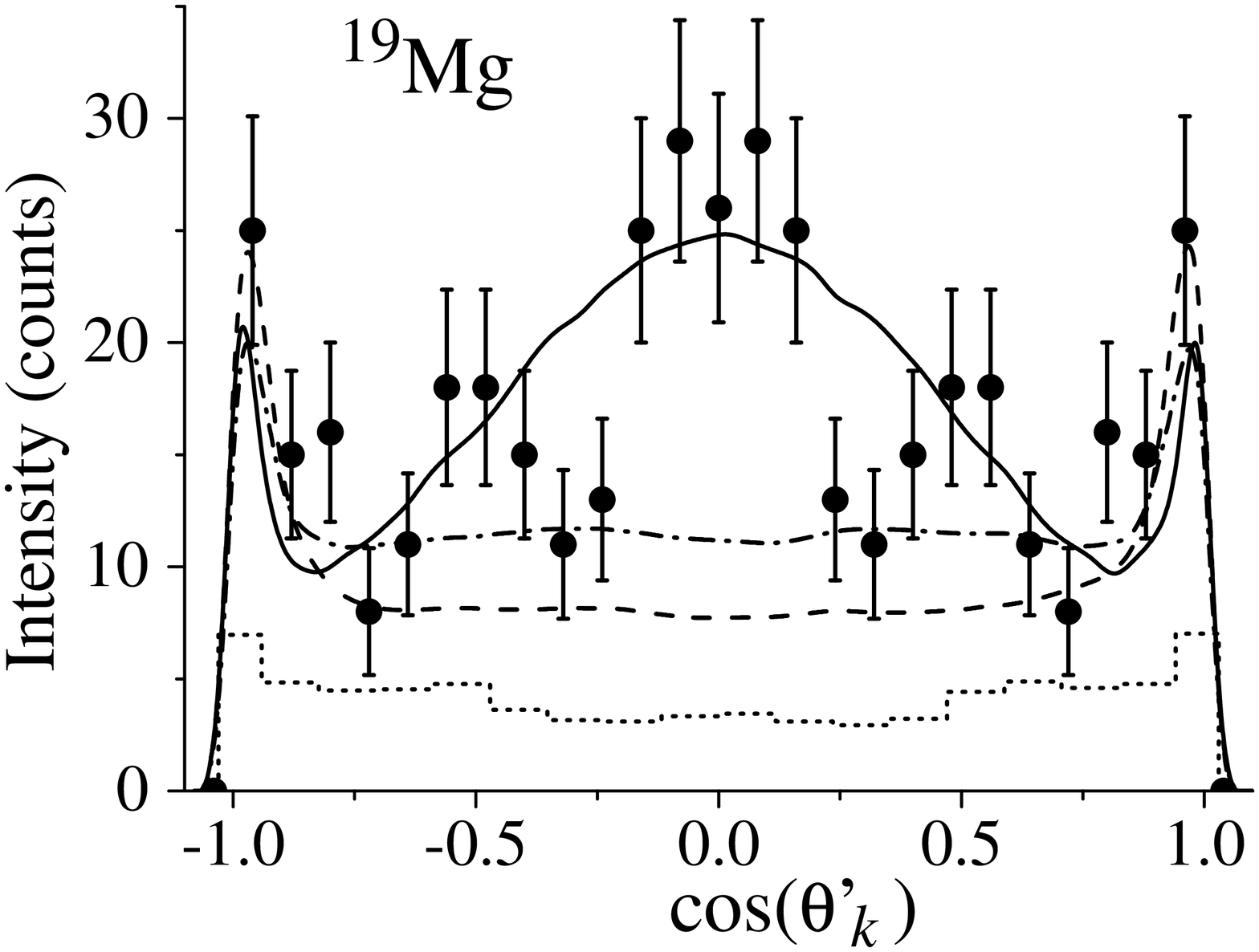}\hspace{-0.2 cm}
\includegraphics[width=0.23\textwidth,angle=0.]{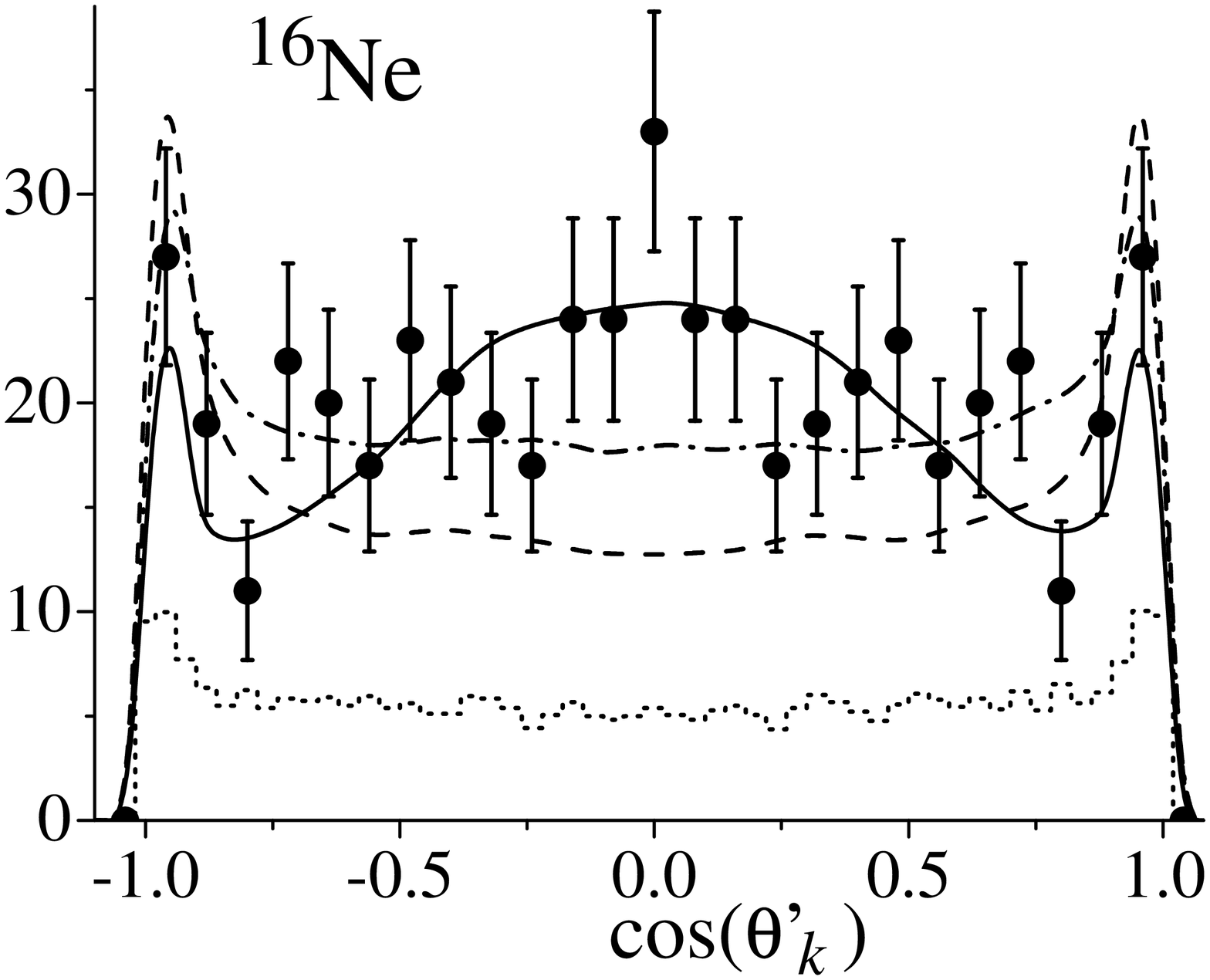}
}
\vspace{-0.2 cm} \caption{
      \label{fig5}
 Three-body correlations from the 2p-decays of the ground states of $^{16}$Ne
 and $^{19}$Mg (full circles with statistical uncertainties)
 obtained from the measured $^{14}$O+p+p and $^{17}$Ne+p+p events by selecting an angle $\theta'_k$ (see text).
 The solid curves are the three-body model calculations (assuming 54\% and 88\% { of $d$-wave
 configuration} in $^{16}$Ne and $^{19}$Mg, respectively).
 The dashed curves are the diproton model predictions, and the dash-dotted curves are the phase-space
 simulations illustrating { the response} to an isotropic 2p-emission in the precursor's rest frames (both simulations are not normalized).
 The dotted curves show the background contributions estimated from all measured HI+p+p events.
}
      \end{figure}

In Fig.~\ref{fig5}, the intensity distributions are displayed as a
function of $\cos(\theta'_k)$.
The angle $\theta'_k$  (see Fig.~\ref{fig1a}(a)) was defined by  a line connecting { the}
two points where two protons hit the same detector and by a vector
joining the { middle between} the 2p hits { and} the point
of a related heavy-ion hit, in analogy with the angle $\theta_k$
shown in Fig.~\protect\ref{fig1a}(a). The typical theoretical
prediction for such a distribution is shown in
Fig.~\ref{fig1a}(c).
The diproton model predicts
flat angular distributions in contrast to the experimental data in
both cases. Only the three-body { model can} reproduce the
characteristic shapes of the observed correlations with the broad
bumps around $\cos(\theta'_k)$=0 (the indicated spikes at
$\cos(\theta'_k)\!\approx\!\pm$1 predicted by all calculations
 are less conclusive).
Such a shape is a manifestation of the ``Coulomb focusing''
efficiently repulsing the fragments from { large regions} of
the momentum space (see Fig.~\ref{fig1a}(a)). These distributions
are weakly sensitive to the assumed structure of the
parent states but are an exclusive feature of the { three-body
model}.

In summary, the measured three-particle correlations from the 2p
decay of the ground states of $^{16}$Ne and $^{19}$Mg are
described quantitatively by the predictions of the tree-body model
\cite{grig03a}, in contrast to the quasi-classical ``diproton''
model which fails to describe our observations. These correlations
are sensitive to the structure of the decaying nucleus. Thus the
comparison between experiment and theory allows one to obtain
spectroscopic information about the parent states. In $^{16}$Ne,
the data are consistent { with strong $s/d$ mixing}
\cite{grig_ne16}. In $^{19}$Mg, the dominating  $d$-shell
configuration is the preferable description which is also
consistent with the lifetime information \cite{mukh_mg19}. The
method of measuring { 2p-decays} in flight by { precisely
tracking all fragments} with microstrip detectors provides new
specific { observables}, thus yielding valuable {
spectroscopic} information on such exotic isotopes. Information
about two-body subsystems, e.g., $^{15}$F, is simultaneously
obtained. Systematic studies of other 2p emitters predicted
theoretically \cite{grig03,grig01a} are foreseen with this novel
technique.

The authors are grateful to M.\ Pohl and his co-workers of the DPNC,
Universite de G\'en\`eve, for developing the microstrip detectors.
We thank in particular E.\ Cortina for the valuable contribution to
this project. We appreciate the help of A.~Bruenle, K.H.~Behr,
W.~Hueller, A.~Kelic, A.~Kiseleva, R.~Raabe and O.~Tarasov during the
 preparations of the experiment.
This work has been supported by contracts
EURONS No. EC-I3 and FPA2003-05958, FPA2006-13807-C02-01 (MEC,
Spain), the INTAS grant 03-54-6545, the Russian RFBR grants 05-02-16404 and 05-02-17535
and the Russian Ministry of Industry and Science grant NS-1885.2003.2.

\end{document}